\documentstyle[eqsecnum,epsfig,aps]{revtex}

\def\Pom{{I\!\!P}}
\def\Reg{{I\!\!R}}

\def\Por{{\tilde{I\!\!P}}}
\begin{document}

\title{
\vspace*{10mm}
RAPIDITY GAPS  AND THE {\sc Phojet} MONTE CARLO\thanks{
Talk presented by J. Ranft at LAFEX International School of
High Energy Physics, Session C: Workshop on diffractive physics
LISHEP98, Rio de Janeiro, Feb.16--20 1998.
The {\sc Phojet} code and write--up can be obtained from
http://lepton.bartol.udel.edu/\~{}eng/phojet.html.
}
}
\author{ F.W.~Bopp}
\address{
{} Universit\"at Siegen, Fachbereich Physik, D--57068 Siegen, Germany, 
e--mail: bopp@physik.uni-siegen.de}
\author{R.~Engel}
\address{ 
University of Delaware, Bartol Research Institute, Newark, DE 19716 USA, 
e--mail: eng@lepton.bartol.udel.edu}
\author{ J.~Ranft}
\address{
{}Universit\"at Siegen, FIGS \& Fachbereich Physik, D--57068 Siegen, Germany, 
e--mail: Johannes.Ranft@cern.ch}
\maketitle
\vspace*{0.5cm}
\centerline{March 1998}
\vspace*{0.5cm}
\vspace*{-7.0cm}
\hspace*{15cm} {Si 98--25}\\
\hspace*{15cm} {BA--98--17}
\vspace*{6.0cm}

\begin{abstract}  
A model for the production of large rapidity gaps being
implemented in the Monte Carlo event generator {\sc Phojet} is
discussed.
In this model, high-mass diffraction dissociation exhibits properties
similar to hadron production in non-diffractive hadronic
collisions at high energies.
Hard diffraction is described using leading-order 
QCD matrix elements together with a parton distribution function 
for the pomeron and pomeron-flux factorization. Since this factorization is
imposed on Born graph level only, unitarity corrections lead to a
non-factorizing flux function.  Rapidity gaps between jets are
obtained by soft color reconnection.
It was previously shown that  this model is able to describe data 
on diffractive hadron production
{}from the CERN-SPS collider and from the HERA lepton-proton collider.
In this work  we focus on the model predictions for rapidity gap events 
 in $p$--$p$ collisions at $\sqrt{s} = 1800$ GeV and 
compare to TEVATRON data. 
\end{abstract}


\section{Introduction}

The {\sc Phojet} event generator is a Monte Carlo implementation of the
two-component Dual Parton Model. This model combines results obtained
within Regge theory, Gribov's reggeon
calculus \cite{Gribov67a-e,Gribov68c-e}
and Abramowski-Gribov-Kancheli (AGK) cutting
rules \cite{Abramovski73-e} with perturbative QCD predictions 
for hard interaction processes 
(see for example \cite{Innocente88,Hahn90,Aurenche92a}, a
review is given in \cite{Capella94a}).

The Dual Parton Model describes high-mass diffractive hadron 
production in terms of enhanced graphs like
the triple-pomeron graph \cite{Kaidalov79}.
As already discussed in \cite{Kaidalov74a}, within this approach,
diffractive processes can be considered as
collisions of a color neutral object, the pomeron, with hadrons,
photons or other pomerons.
However, it is important to note that the pomeron cannot be considered as
an ordinary hadron. It is only a theoretical object 
providing an effective description of the important 
degrees of freedom of a certain sum of Feynman diagrams in Regge limit (e.g.\
the available c.m. energy is large compared to the momentum transfer
characterizing the scattering process).
In this sense, pomeron-hadron
or pomeron-pomeron interactions can only be discussed in the framework of
collisions of other particles like hadrons or photons.

Experimental data support the interpretation of
diffractive particle production in terms of pomeron-hadron or
pomeron-pomeron collisions. It was found that
high-mass diffraction
dissociation exhibits similar features as non-diffractive hadron
production where the mass of the diffractively produced system
corresponds to the collision energy in non-diffractive interactions
\cite{Goulianos83,Bernard86a}.
Furthermore, it is well known that, in order to obtain a reasonable 
Monte Carlo description of 
non-diffractive hadron production, multiple partonic interactions 
between projectile and target are needed at high energies
\cite{Aurenche84a,Sjostrand87b,Aurenche92a}.
Given the striking similarities between diffractive and
non-diffractive multiparticle production one may expect that multiple
soft as well as hard interactions may also play an important role in 
pomeron-hadron/photon/pomeron interactions.

Another consequence of the triple-pomeron graph interpretation is
an energy-dependent normalization of the pomeron flux. Experimentally,
the pomeron flux in target diffraction dissociation is given by the 
cross section for rapidity gap
events divided by the corresponding pomeron-target cross section.
However, since at high energies additional
projectile-target interactions are likely to fill the rapidity
gap with hadrons, the experimentally observable
pomeron flux is smaller
than the flux implied by the triple-pomeron graph. Furthermore, it 
may depend not only on projectile but also on target
properties.

However, in the case of a large rapidity gap between the jets
in two-jet events, as observed at TEVATRON
\cite{Abachi96b,Abe98a}, at least one pomeron has a 
large virtuality and standard Regge phenomenology cannot be
applied. Therefore, one has to introduce a new kind of process like
soft color reconnection (SCR)
\cite{Buchmuller95a,Buchmuller95b,Ingelman95a,Amundson96a,Eboli97a}
or perturbative gluon-ladder exchange \cite{Mueller92a,DelDuca93a}.

The {\sc Phojet} Monte Carlo is a first attempt to built a model 
which accounts for both effects, multiple soft and hard interactions
between the constituents of the projectile and 
target as well as multiple interactions in the pomeron-projectile,
pomeron-target, and pomeron-pomeron scattering subprocesses.
The model includes also SCR in hard scattering processes similar to
\cite{Eboli97a}.



\section{The Model\label{the-model}}
%
%
The realization of the DPM in {\sc Phojet} \cite{Engel95a,Engel95d}
with a hard and a soft
component is similar to the event generator {\sc Dtujet}
\cite{Aurenche92a,Bopp94a} for $p$--$p$ and $\bar p$--$p$ collisions. 
Interactions of hadrons are described in terms of
reggeon ($\Reg$) and pomeron ($\Pom$) exchanges. 
The pomeron exchange is artificially
subdivided into  {\it soft}
processes and processes with at least one large momentum
transfer ({\it hard} processes).
This allows us to use the predictive power of the QCD-improved Parton
Model with lowest-order QCD matrix elements \cite{Combridge77,Duke82a} and
parton distribution functions (PDFs).
Practically, soft and hard processes are distinguished by applying
a transverse momentum cutoff $p_\perp^{\mbox{\scriptsize cutoff}}$ 
of about 3 GeV/$c$ to the scattered partons. The pomeron is considered as a 
two-component object with
the Born graph cross section for pomeron exchange given by 
the sum of hard and soft cross sections.

\begin{figure}[!htb]
\centerline{\epsfig{file=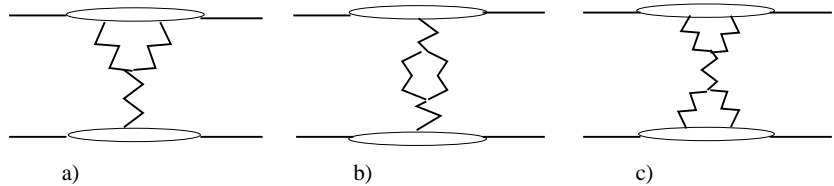,width=11cm}}
\medskip
\caption{Enhanced pomeron exchange graphs considered in the model: a)
triple-pomeron, b) loop-pomeron, and c) double-pomeron graphs. The
zig-zag lines represent pomeron propagators.
\label{enh-gra}
}
\end{figure}
High-mass diffraction dissociation is described in the following way.
In order to get an efficient parametrization of Born graph cross
sections describing
diffraction within Gribov's reggeon calculus, we calculate the triple-,
loop- and double-pomeron graphs shown in Fig.~\ref{enh-gra}
using a renormalized pomeron intercept
$\alpha_\Por = 1+\Delta_\Por = 1.08$. The corresponding formulae are
given in \cite{Engel96e}.

Shadowing corrections are approximated by unitarizing
the enhanced graphs together with the leading one-pomeron
exchange (including soft and hard contributions)
in a two-channel eikonal model \cite{Aurenche92a,Engel95a,Engel95d}.

In case of diffractive multiparticle production we have to consider,
in addition to the shadowing contribution from multiple pomeron exchange
between projectile and target, also rescattering effects in 
pomeron-hadron and pomeron-pomeron interactions.
For the cross section calculation the introduction of a
renormalized pomeron trajectory takes this effect into account.
However, for the calculation of particle production, a model for 
the hadronic final
states which correspond to the unitarity cut of such a renormalized
pomeron propagator is needed.
{}Following Refs.~\cite{Cardy74a,Kaidalov86d} we assume that  the 
pomeron-pomeron coupling can be described by the formation of an intermediate
hadronic system $h^\star$ the pomerons couple to. Assuming that
this intermediate hadronic system has properties similar to a pion,
the $n$-$m$ pomeron coupling $g_{n-m}$ reads \cite{Kaidalov86d}
\begin{equation}
g_{n-m} = G \prod_{i=1}^{n+m-2} g_{h^\star\Pom}
\label{n-m-coupling}
\end{equation}
with $g_{h^\star\Pom} = g_{\pi\Pom}$ being the pomeron-pion coupling.
$G$ is a scheme-dependent constant.
Hence, pomeron-hadron and pomeron-pomeron scattering 
should exhibit features similar 
to pion-hadron and pion-pion scattering, respectively.



To introduce hard interactions in diffraction dissociation, the 
impact parameter amplitude of the exchanged
(renormalized) 
pomerons in pomeron--hadron and pomeron--pomeron scattering is again
interpreted as the eikonalized amplitude of soft and hard interactions
\begin{equation}
a_{A\Pom}(M_{\rm D},\vec B) \approx 
\frac{i}{2}\ G\ \left\{ 1 - \exp\left[-\chi_{\rm S}^{\rm diff}(M_{\rm
D},\vec B)
-\chi_{\rm H}^{\rm diff}(M_{\rm D},\vec B) \right] \right\}\ .
\label{two-comp-diff0}
\end{equation}
The diffractive eikonal functions read
\begin{eqnarray}
\chi_{\rm S}^{\rm diff}(M_{\rm D},\vec B) &=& 
\frac{g_{A\Pom}^0
g_{h^\star\Pom}^0 (M_D^2/s_0)^{\Delta_\Pom}}{8 \pi
b_\Pom(M_D^2)}
\exp\left( -\frac{\vec{B}^2}{4 b_\Pom(M_D^2)}\right)
\label{two-comp-diff1}
\\
\chi_{\rm H}^{\rm diff}(M_{\rm D},\vec B) &=&
\frac{\sigma_{\rm hard}^{A\Pom}(M_D^2)}{8 \pi
b_{\rm h,diff}}\exp\left( -\frac{\vec{B}^2}{4 b_{\rm h,diff}}\right),
\label{two-comp-diff2}
\end{eqnarray}
where $\sigma_{\rm hard}^{A\Pom}$ is the parton model cross section for
hard pomeron--$A$ scattering ($A$ can be a hadron, photon or pomeron).
In all calculations the pomeron PDFs proposed by 
Capella, Kaidalov, Merino, and Tran (CKMT) \cite{Capella95a,Capella96a} 
with a hard gluon component are used.


To estimate the sensitivity of the model results to non-factorizing
coherent pomeron 
contributions as proposed in \cite{Collins93,Collins95a}, we 
use optionally also a toy model with a
direct pomeron-quark coupling \cite{Kniehl94}. 
In this case, the pomeron is treated similar
to a photon having a flavor independent, unknown quark coupling $\lambda$.
The corresponding matrix elements are given in \cite{Engel97d}.

%


\section{Soft color reconnection}

Both the CDF and D0 collaborations have found dijet production
by color--singlet exchange \cite{Abachi96b,Abe98a}.
These Jet--gap--Jet
(JgJ) events are not due to traditional diffractive processes.
The two jets separated by a rapi\-dity gap are in polar angle back--to--back
correlated. Certainly in double--diffractive events describing
the diffractively produced systems on both sides of the gap by
pomeron--hadron scattering, we would also find jets,
but these jets would not be back--to--back
correlated. Therefore, we have to consider these events as mainly due
to a new mechanism  of  hard pomeron exchange.

To describe these events within the {\sc Phojet} Monte Carlo, we
introduce SCR between hard scattered partons in nondiffractive
events following Eboli, Gregores and Halzen \cite{Eboli97a}.
This mechanism is quite similar to the the soft color interaction mechanism
described by Ingelman \cite{Ingelman98}.
We use the following SCR probabilities \cite{Eboli97a} in {\sc Phojet}
\begin{equation}
{}F_{qq}:F_{qg}:F_{gg}=\frac{1}{9}:\frac{1}{24}:\frac{1}{64}.
\end{equation}

The simplest hard q--q event, where SCR leads to a rapidity gap between
two jets is an event with just one single hard valence--quark --
valence--quark scattering. In normal events in the Dual Parton
Model we get two color strings 
each being stretched between one scattered quark and the 
diquark of the other
hadron, no rapidity gap is present in such events.
In events with SCR we get a color reconnection caused by
the exchange of soft gluons, now the color strings connect the hard
scattered quark and the diquark of the same hadron, these are
events with a rapidity gap.

The simplest hard g--g event where SCR leads to a rapidity gap
between two jets is an event with just one hard g--g
scattering. In normal events we get again two color strings
connecting the (soft) valence quark of one hadron via the hard
scattered gluon to the (soft) diquark of the other hadron. In
events with SCR the color strings are stretched from the (soft) 
valence quark
of one hadron via one hard scattered gluon to the (soft) diquark
of the same hadron. Such events might have a rapidity gap.

In most events we have multiple soft and hard interactions, even
if a rapidity gap appears in one of the multiple collisions, the
gap might be filled by hadrons resulting from the other
collisions. The Monte Carlo simulation of complete events 
incorporates this effect,
in this way {\sc Phojet} accounts already for the gap 
survival probability \cite{Bjorken93a,Bopp97a}.


\section{Previous comparisons with hadron--hadron and
Photon--hadron data\label{comparison-diff}}


\subsection{Diffractive cross sections}

Studying diffractive cross sections is not the primary concern
of this paper. Results on diffractive cross sections were
already presented  using the {\sc Dtujet} model in
Refs.~\cite{Aurenche92a,Engel92a,Bopp94a} and using the present {\sc
Phojet} model in Refs.~\cite{Engel95a,Engel96e}, we include
updated results for these cross sections here.

\begin{figure}[thb] \centering
\begin{center}
\unitlength1mm
\begin{picture}(135,62)
\put(0,0){\epsfig{figure=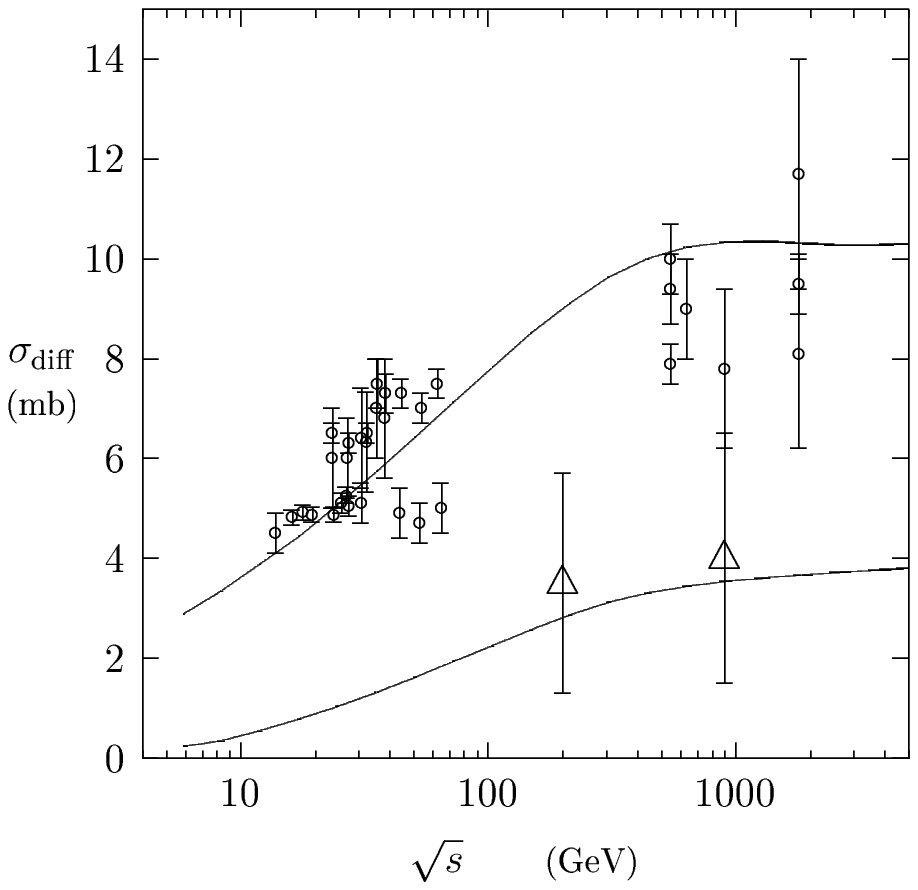,width=6.0cm}}
\put(2,3){a)}
\put(65,0){\epsfig{figure=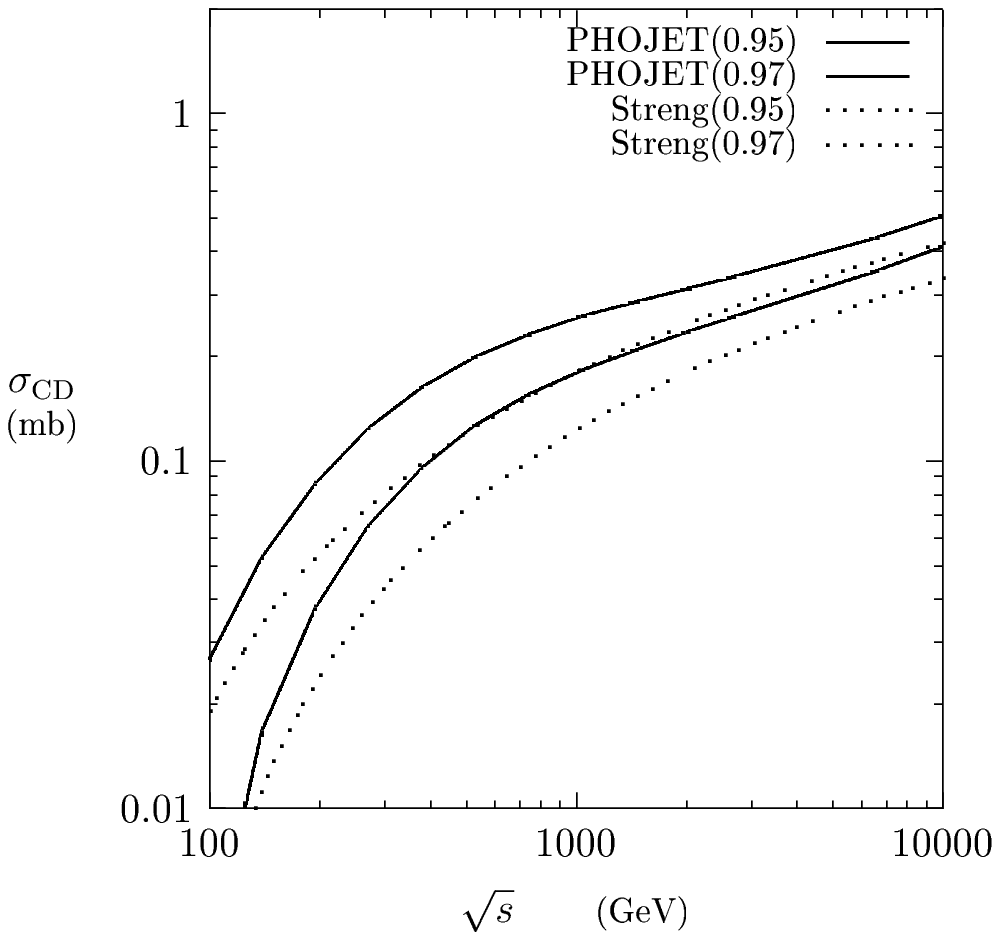,width=6.2cm}}
\put(67,3){b)}
\end{picture}
\end{center}
\vspace*{-3mm}
\medskip
\caption{(a)
Single and double diffractive $p-\bar p$ cross sections as a
{}function of the center of mass energy $\protect\sqrt s$. Model results
are compared to data on single diffractive cross sections
\protect\cite{Chapman74,Schamberger75,Albrow76,Armitage82,%
Ansorge86,Robinson89,Amos90a,Amos93a,Abe94c}. 
In addition, some experimental
estimates for the cross section on double diffraction dissociation
\protect\cite{Ansorge86,Robinson89} are shown (triangles).
(b) The energy dependence of the central diffraction
cross section. We compare the cross section as obtained from
\protect{\sc Phojet} with unitarization using a supercritical
pomeron with the cross section obtained by Streng
\protect\cite{Streng86a} without unitarization and with a critical
pomeron. Both cross sections are for the same two kinematic cuts:
$M_{\rm CD}>2$ GeV/c${}^2$ and Feynman-$x$ of the scattered hadron
$x_F >0.95$ (upper curves) and $0.97$ (lower curves).
\label{ppdif}
}
\end{figure}

In Fig.~\ref{ppdif}.a data on single diffractive cross sections
\cite{Chapman74,Schamberger75,Albrow76,Armitage82,%
Ansorge86,Robinson89,Amos90a,Amos93a,Abe94c}
are compared with our model results ($M_{\rm D}^2 < 0.05 s$). 
It is to be noted that the
data on single diffractive cross sections at collider energies
are subject to large uncertainties. Nevertheless the rise of the cross
section from ISR energies to the energies of the CERN and
{}FERMILAB colliders is less steep than expected from the Born level
expression which is the triple pomeron formula.
However, within our model a renormalized pomeron flux as proposed in
\cite{Goulianos95a} is not needed. There are two reasons this:

(i) The eikonal unitarization procedure in the model
suppresses the rapidity gap survival probability. This effect is well
known (see for example \cite{Capella76,Gotsman95a}) and can be tested
directly comparing diffraction dissociation in deep inelastic scattering
and photoproduction at HERA \cite{Bopp97a}.

(ii) The graph for double-pomeron scattering has cuts which correspond to
single diffraction dissociation. However, due
to the negative sign of these contributions the 
diffractive cross section is significantly reduced at high energies as
compared to a model with an eikonalized triple pomeron graph only
\cite{PhD-RE}.

In Fig.~\ref{ppdif}.b we compare as function of the energy 
the central diffraction cross
sections in proton-proton collisions, 
which we obtain from {\sc Phojet} with the cross
section calculated by Streng \cite{Streng86a}. 
In {\sc Phojet} we use a supercritical pomeron with
$\Delta_{\Por}$ = 0.08 whereas Streng \cite{Streng86a}
uses a critical Pomeron with $\Delta_{\Pom}$ = 0.
Note that also the double-pomeron cross section 
grows in Born approximation
with $s$ like 
$\sim s^{2\Delta_\Por}$. This rapid increase is damped 
in {\sc Phojet} by the unitarization procedure. At high energies, 
contributions from multiple interactions become important. 
The  rapidity gaps are filled with hadrons due to 
inelastic rescattering and the cross section for central diffraction
gets strongly reduced. In contrast, Streng
calculates only the Born term cross section.
{}Figure~\ref{ppdif}.b
illustrates the  differences obtained using 
different theoretical methods.
We stress, both methods use the measured
single diffractive cross sections to extract the triple-pomeron
coupling.


\subsection{Diffractive hadron- and jetproduction}

There are some experiments on diffractive particle production
at colliders, which we have studied previously using {\sc Phojet}
\cite{Engel97d,Engel95c}. 
Generally, we have reached a good agreement. We do not present
these comparisons again here.

Among others, the following experiments 
have studied hadron production in single diffraction
dissociation at the CERN--SPS and DESY--HERA and colliders:
\begin{enumerate}
\item
The UA--4 Collaboration \cite{Bozzo84b,Bernard86a,Bernard87b}
measured pseudorapidity distributions of charged hadron
production for different masses of the diffractive system. We
have already twice compared earlier versions of the 
Dual Parton Model \cite{Ranft87c,Roesler93} 
to this data.  New in the present model is 
hard diffraction and multiple interactions in the diffractive hadron
production, therefore we have again compared to this data
and we find a reasonable agreement \cite{Engel97d}. In the model,
multiple interactions and minijets lead to a rising rapidity plateau in
pomeron--proton collisions in a similar way as observed 
in hadron--hadron collisions. 
\item
Hard diffractive proton--antiproton interactions were
investigated by the UA--8 Collaboration \cite{Brandt92}. In this
experiment the existence of a hard component of diffraction was
demonstrated for the first time. Because of the importance of
these findings, we compared them already in 
\cite{Engel95c} to
our model and found the model  to be consistent with this experiment. 
Therefore we will not repeat this comparison here.
\item
Results on single photon diffraction dissociation
and in particular hard single
diffraction were presented by both experiments at the HERA
electron--proton collider
\cite{Ahmed95a,Aid95b,Derrick95a,Derrick95h,Derrick95i}.
The ZEUS Collaboration \cite{Derrick95h} 
has presented differential and integrated
jet pseudorapidity cross sections for jets with $E^{\rm jet}_{\perp} >$ 8
GeV. The absolute normalization of these data is given. This
allows a severe check of the model. In \cite{Engel97d} 
 we have compared the differential  jet
pseudorapidity cross sections from ZEUS 
\cite{Derrick95h} 
to the model. 

\end{enumerate}

\section{Comparing hadron production in diffractive processes to
non-diffractive particle production in $p$--$p$  and $\gamma$--$\gamma$
reactions\label{comparison-channel}}

\begin{figure}[thb] \centering
\begin{center}
\unitlength1mm
\begin{picture}(135,62)
\put(0,0){\epsfig{figure=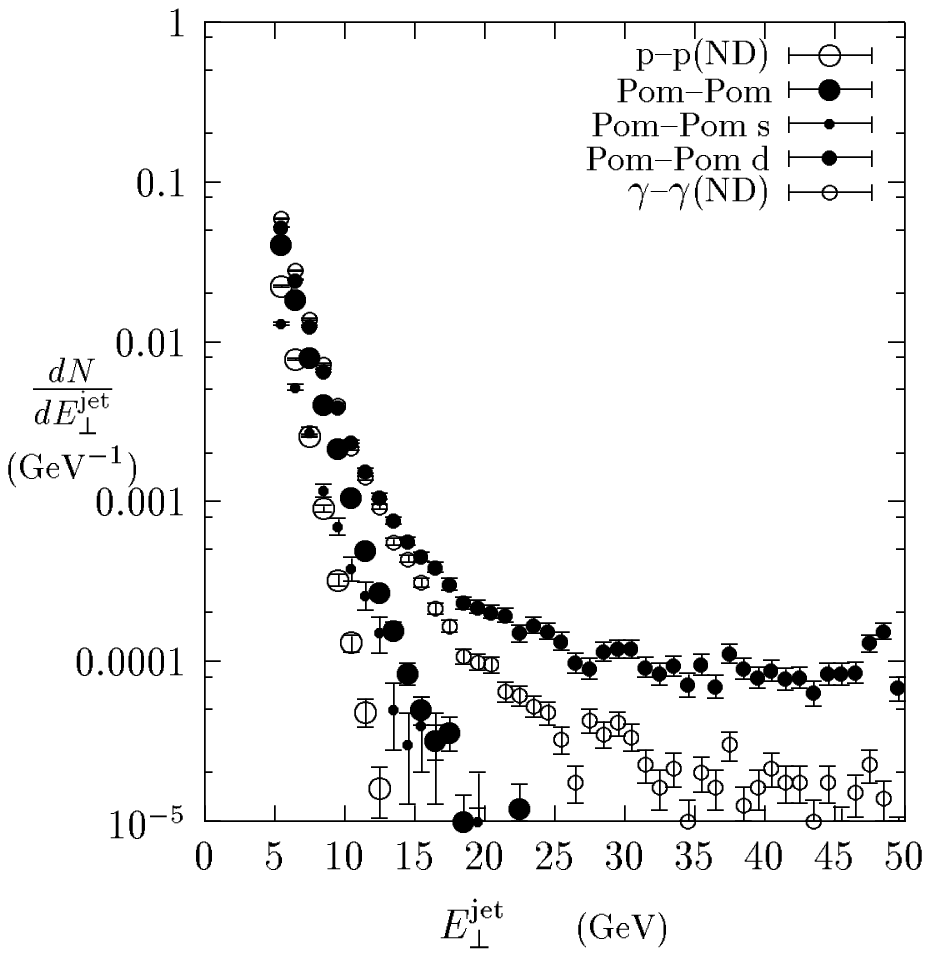,width=6.0cm}}
\put(2,3){a)}
\put(65,0){\epsfig{figure=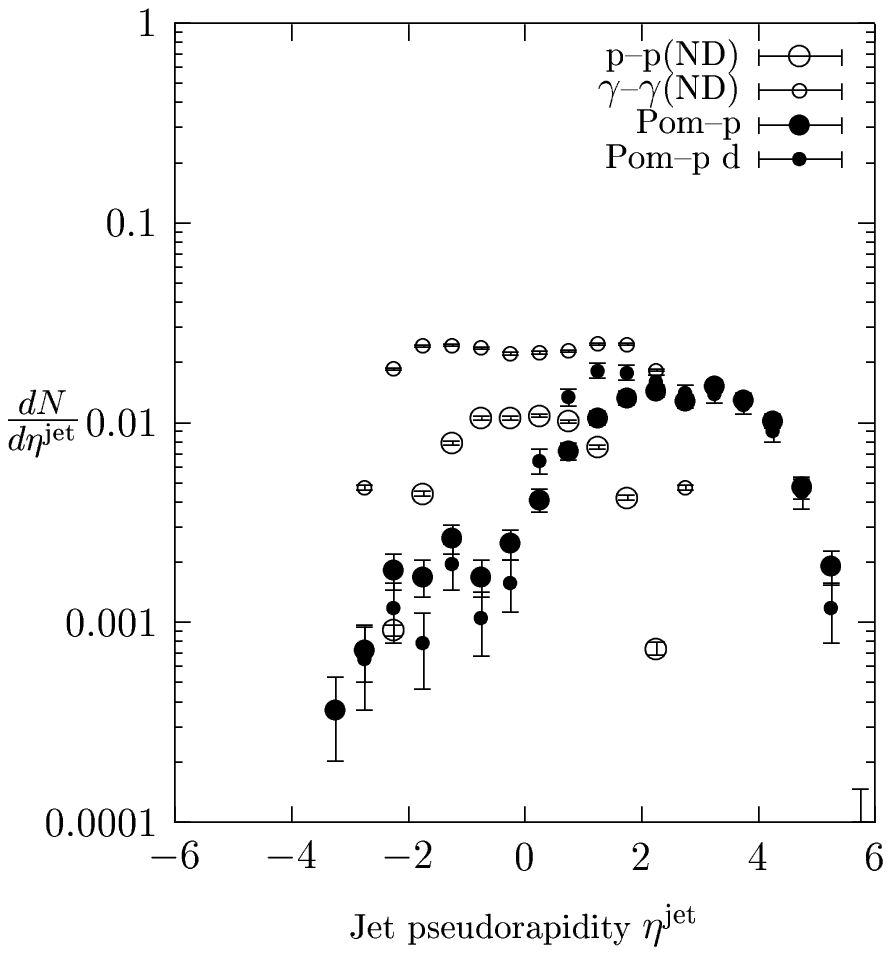,width=6.0cm}}
\put(67,3){b)}
\end{picture}
\end{center}
\vspace*{-3mm}
\medskip
\caption{(a)
Jet transverse energy  distributions in non-diffractive $p$--$p$ and
$\gamma$--$\gamma$ 
 collisions compared with the jet transverse
energy  distribution in 
central  diffraction (pomeron--pomeron 
 collisions). For the latter channel we give the distributions
 separately for the full model, the model without multiple
 interactions (s) and the model with a direct pomeron coupling
 (d). The distributions
were generated with \protect{\sc Phojet}, 
the c.m.\ energy / diffractive
mass is 100 GeV in all cases.
(b)
Jet pseudorapidity  distributions in non-diffractive $p$--$p$ and
$\gamma$--$\gamma$ collisions compared with the jet pseudorapidity
  distribution in 
 single diffraction (pomeron--$p$ 
scattering). The distributions
were generated with \protect{\sc Phojet}, again the c.m.\ energy /
diffractive mass is 100 GeV in
all  cases, but the pseudorapidities  in the collisions with
pomerons given refer to the $\protect\sqrt
s$ = 2 
TeV $p$--$p$  collisions used to generate the
diffractive events.
\label{pt100jpopo}
}
\end{figure}
%
%

In   Sections II  we have already pointed out, that our model for
particle production in pomeron--hadron/photon collisions and
pomeron--pomeron collisions has the same structure characterized
by multiple soft collisions and multiple minijets like models
{}for hadron production in non-diffractive hadron--hadron collisions. 
Therefore,
again we expect the main differences in comparison to other channels in
the hard component due to the differences between the pomeron
and hadron structure functions and
due to the existence or nonexistence of a direct
pomeron--quark coupling. 

The differences
in the parton structure functions of protons, photons and
pomerons lead to quite different energy dependences of the hard
cross sections. In all processes where pomerons are involved,
single diffraction and central diffraction, hard
processes become important already at lower energies. For
pomeron--pomeron scattering at low energy the hard cross section
is about a factor 100 bigger than that of $p$--$\bar p$ collisions. At
high energies the opposite happens, the hard cross sections in
all processes where pomerons are involved rise less steeply with
the energy than in purely hadronic or photonic processes. The
reason for this is the different low-$x$ behavior of the
parametrization of the structure functions used. However, nothing
is known at present from experiment 
about the low-$x$ behavior of the pomeron
structure function.

In Fig.~\ref{pt100jpopo}.a we compare jet transverse
energy distributions  in $p$--$p$ and
$\gamma$--$\gamma$ collisions with the ones in 
$\Pom$--$\Pom$ collisions.  In the  channels with pomerons
we present again the distributions according to our full model, 
according to the model without multiple interactions and the
model with a direct pomeron--quark coupling. 
In  all non-diffractive collisions we have  $\sqrt s$ =
100 GeV and the diffractive events are generated in $\sqrt s $ = 2
TeV collisions with $M_D = 100$ GeV/$c^2$. 
The differences in the jet transverse energy
distributions between the channels
are as to be expected 
more pronounced than in the hadron $p_{\perp}$ distributions.
We observe an important reduction in the jet distributions in
the model without multiple interactions. The effect of the
direct pomeron coupling is as dramatic as the effect due to the
direct photon coupling. The $E_{\perp}$ distributions in the
$\Pom$--$\gamma$ and $\Pom$--$\Pom$ channels extend up to the
kinematic boundary. In the latter two cases 
as in the case of $\gamma$--$\gamma$ 
collisions the entries at large $E_{\perp}$ come
only from direct processes.

In Fig.~\ref{pt100jpopo}.b we compare jet 
pseudorapidity distributions in $p$--$p$,
$\gamma$--$\gamma$ and $\Pom$--$p$, again, all collisions at $\sqrt s$ =
 100 GeV with the diffractive events generated in $\sqrt s $ = 2
TeV collisions.
{}For the jets we observe  substantial
differences in the shape of the pseudorapidity distributions.


\section{Single diffraction and central diffraction at
TEVATRON}

\begin{figure}[thb] \centering
\epsfig{figure=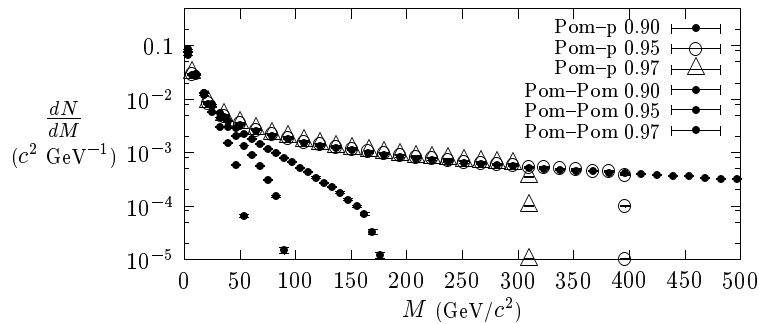,width=10.0cm}
\medskip
\caption{
Distribution of the diffractive mass in single diffraction dissociation
(pomeron--proton) and central diffraction (pomeron--pomeron) at
TEVATRON with $\sqrt s = 1.8$ TeV for three different cuts of
the Feynman-$x$ of the diffractive nucleons.
\label{fdndm}
}
\end{figure}

\begin{figure}[thb] \centering
\begin{center}
\unitlength1mm
\begin{picture}(135,62)
\put(0,0){\epsfig{figure=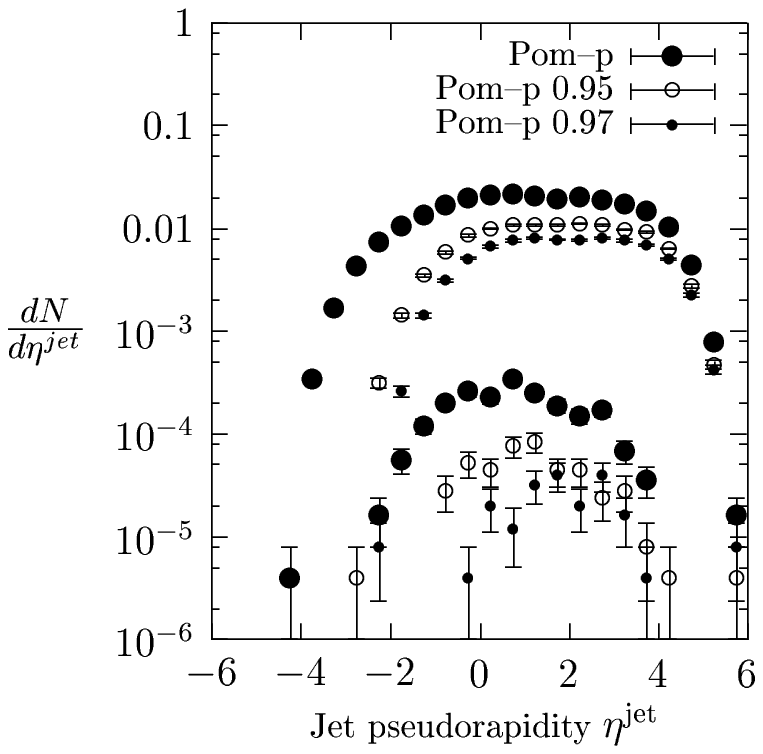,width=6.0cm}}
\put(2,3){a)}
\put(65,0){\epsfig{figure=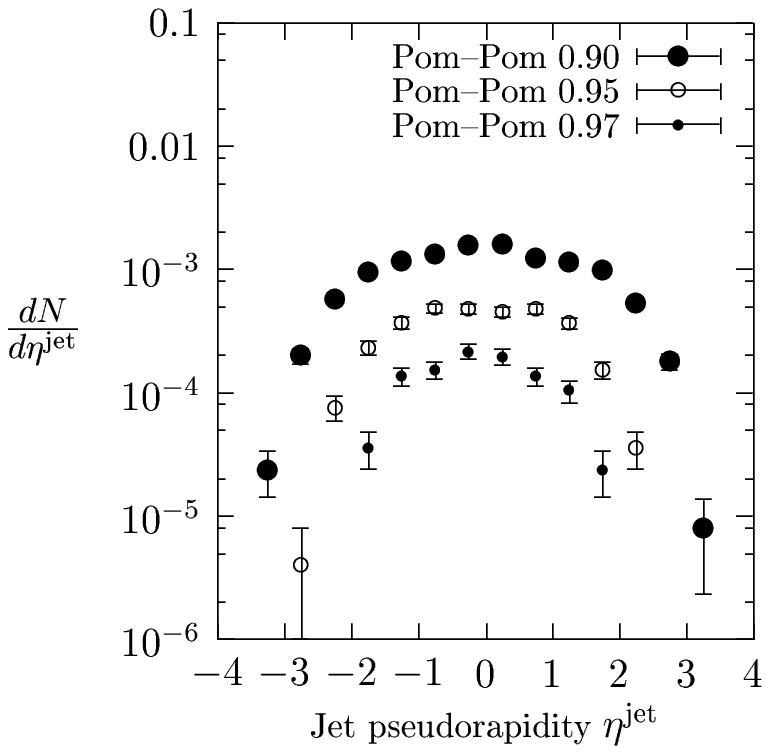,width=6.0cm}}
\put(67,3){b)}
\end{picture}
\end{center}
\vspace*{-3mm}
\medskip
\caption{(a)
Pseudorapidity distribution of jets with $E_{\perp}$ larger
than 5 GeV  and 15 GeV 
in (one side) single diffraction (Pom--p) 
 at TEVATRON for three different cuts of the Feynman-$x$ of
 the diffractive nucleon.
The upper curves with the same plotting symbol are generally 
{}for $E_{\perp}$ = 5 GeV, the lower curves are for $E_{\perp}$ = 15 GeV.
(b)
Pseudorapidity distribution of jets with $E_{\perp}$ larger
than 5 GeV 
in  central diffraction (Pom--Pom) 
 at TEVATRON for three different cuts of the Feynman-$x$ of
 the diffractive nucleons.
\label{fdndetaj}
}
\end{figure}

In Figs.~\ref{fdndm} to \ref{fdndetaj}.b we present some cross
sections calculated using {\sc Phojet} at TEVATRON energy. The
distributions are mass distributions in single and central
diffraction (Fig.~\ref{fdndm}) and jet pseudorapidity distributions
in single and central diffraction using  $E_{\perp}$
thresholds of 5 and 15 GeV (Fig.~\ref{fdndetaj}.a and .b).
In all Figs. we give the plots for three different cuts for the
{}Feynman-$x$ of the diffractive nucleons $x_F>$  0.9, 0.95 and
0.97. It is obvious, that all distributions and cross sections
depend strongly on these cuts.

One of the results obtained by the D0 Collaboration
is the ratio of double--pomeron exchange (DPE)\footnote{
In \protect\cite{Albrow97a} the term
double--pomeron exchange is used instead of central diffraction.}
to non--diffractive (ND) dijet events \cite{Albrow97a}:
\begin{equation}
\left(\frac{\sigma({\rm DPE})}{\sigma({\rm ND})}\right)_{E_{\perp}^{\rm
jet}>15 {\rm GeV}} 
\approx 10^{-6}
\end{equation}
Within the {\sc Phojet} model one gets the following cross sections:\\
\hspace*{1cm}Non-diffractive interactions (ND): 
$\sigma ({\rm ND}) = $ 45.2 mb,\\
\hspace*{1cm}Both-side single diffraction dissociation (SD): 
$\sigma ({\rm SD}) =$ 11.2 mb,\\
\hspace*{1cm}Central diffraction (CD): $\sigma ({\rm CD}) =$ 0.64 mb.\\
{}From these cross sections together with Figs.\ like \ref{fdndetaj}  
 we get, always calculated for
$E_{\perp}^{\rm jet}$ larger than 15 GeV:\\
\hspace*{1cm}(CD)/(ND)$\approx 2 \times 10^{-6}$,\\
\hspace*{1cm}(SD)/(ND)$\approx 4 \times 10^{-3}$,\\
\hspace*{1cm}(CD)/(SD)$\approx 0.5 \times 10^{-3}$.\\
Despite the fact that no experimental acceptance has been considered for
these {\sc Phojet} results it is interesting to find the (CD)/(ND)
ratio so close to the D0 value given above.
%
%
%
%
%
\section{Diffractive dijet production at TEVATRON}

Data on dijet production in single diffraction dissociation using a rapidity
gap trigger were published by the CDF Collaboration \cite{Abe97a}.
Same side ($\eta^{{\rm jet}1} \times \eta^{{\rm jet}2} > 0$) 
dijets were selected with
$E^{\rm jet}_{\perp}>$  20 GeV in the jet pseudorapidity window 1.8 
$<|\eta^{\rm jet}|<$  3.5. The gap trigger did demand no charged
hadrons in the range 3.2  $<|\eta |<$  5.9 opposite to the jets
and no calorimeter hit above 1.5 GeV in the range 2.4  $<|\eta |<$
 4.2 opposite to the jets.
The ratio of dijet events with gap (JJg) to dijets without
gap (JJ) was found to be 
\begin{equation}
R_{\rm JJg-CDF} = \frac{\rm (JJg)}{\rm (JJ)} = (0.75 \pm 0.05 \pm 0.09)
\%\ .
\end{equation}

Using {\sc Phojet} 
we got so far good statistics only for $E^{\rm jet}_{\perp}>$ 
10 GeV when using the CDF pseudorapidity restrictions. We obtained the
cross sections $\sigma_{\rm JJ}$ = 50.4 $\mu$b and $\sigma_{\rm JJg}$ =
0.107 $\mu$b. This gives the ratio $R_{\rm JJg- PHOJET}$ = 0.21$\%$.
There are two possible reasons for this ratio being smaller
than the one found by CDF: (i) the different $E_{\perp}$ cut and
(ii) The CKMT pomeron structure functions \cite{Capella95a,Capella96a}
used in the calculation might not contain enough hard gluons.

\begin{figure}[thb] \centering
\begin{center}
\unitlength1mm
\begin{picture}(135,62)
\put(0,0){\epsfig{figure=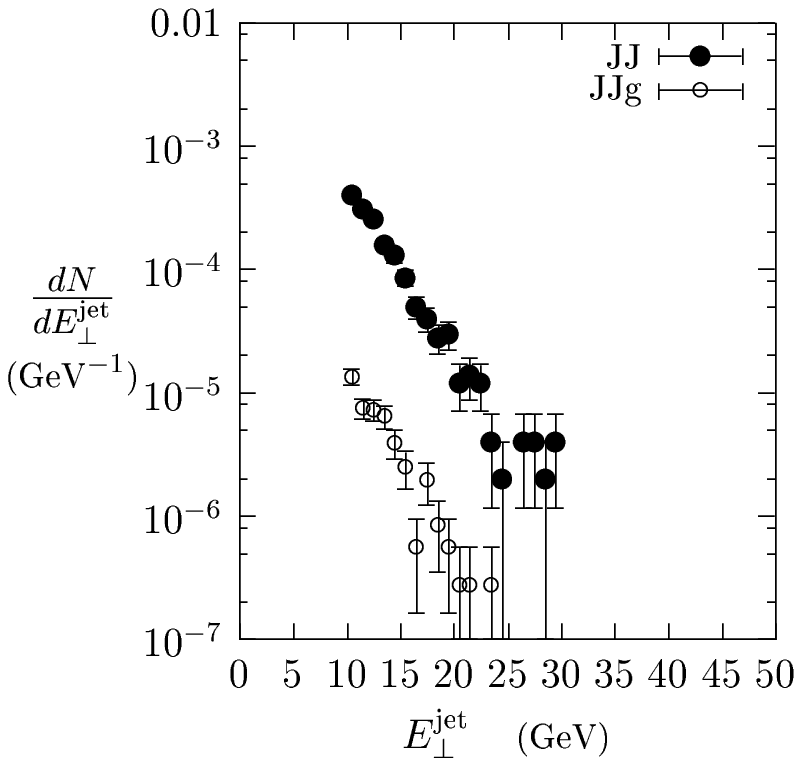,width=6.0cm}}
\put(2,3){a)}
\put(65,0){\epsfig{figure=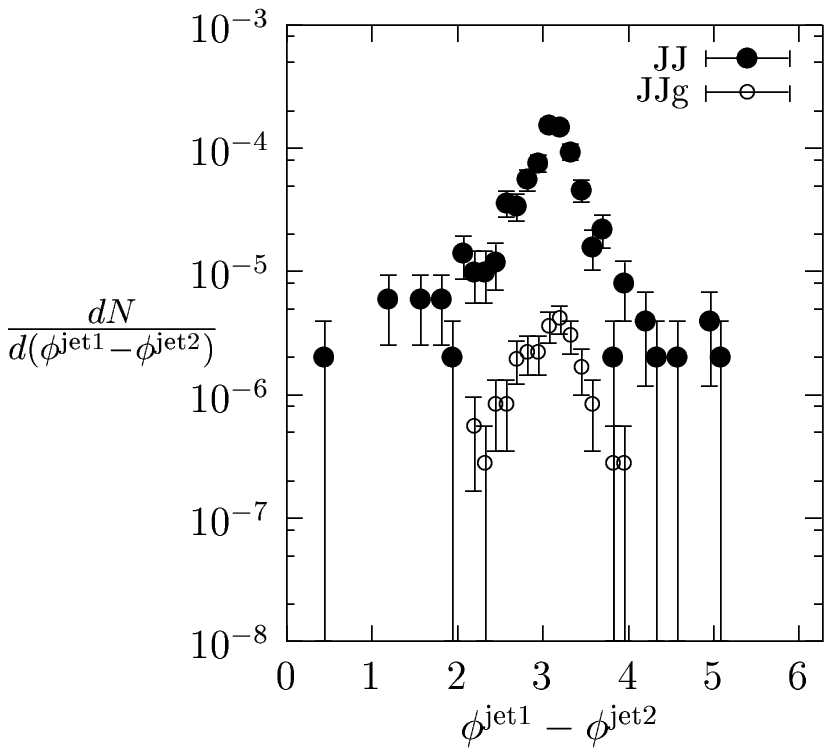,width=6.0cm}}
\put(67,3){b)}
\end{picture}
\end{center}
\vspace*{-3mm}
\medskip
\caption{(a)
$E^{\rm jet}_{\perp}$ distributions in  JJ and JJg events obtained in
{\sc Phojet} using the CDF triggers.
(b)
$\phi^{{\rm jet}1}$ - $\phi^{{\rm jet}2}$ distributions in JJ and JJg events
obtained with {\sc Phojet} using the CDF trigger.
\label{jjgdndetj}
}
\end{figure}
%
In Fig.\ref{jjgdndetj}.a we present $E_{\perp}^{\rm jet}$
distributions calculated from {\sc Phojet} for the JJ and JJG events.
Within the statistics of the Monte Carlo calculation both
distributions seem to have the same shape.

In Fig.\ref{jjgdndetj}.b we present $\phi^{{\rm jet}1}-\phi^{{\rm jet}2}$
distributions for the JJ and JJg events. Again within the statistics
both distributions seem to be quite similar. However, in the JJ
events we find more often additional jet-pairs than in the JJg events.
Therefore we would expect a more narrow correlation of the two
jets in the JJg events.

%

\section{Dijet production by color--singlet exchange at TEVATRON}

We will refer here only to the data on dijet production
 by color--singlet exchange published by the CDF and D0
Collaborations \cite{Abachi96b,Abe98a}. More preliminary data have been
  presented at this meeting. 
  
  D0 \cite{Abachi96b} finds opposite
  side ($\eta^{{\rm jet}1}\times \eta^{{\rm jet}2} < 0$) dijets 
  with $E^{\rm jet}_{\perp}>$ 30 GeV and $|\eta^{\rm jet}|>$ 2. The
  pseudorapidity gap is at $|\eta|<$ 1.3. The fraction of JgJ
  events is found to be 
\begin{equation}
R_{\rm JgJ-D0} = \frac{\rm (JgJ)}{\rm (JJ)} 
= (1.07 \pm 0.10 ^{+ 0.25}_{-0.13}) \%
\end{equation}

CDF \cite{Abe98a} uses opposite side jets with $E^{\rm jet}_{\perp}>$
20 GeV and $3.5 >|\eta^{\rm jet}|>$ 1.8 with a gap at $|\eta|<$ 1.0.
The gap fraction is found to be 
 \begin{equation}
 R_{\rm JgJ-CDF} = \frac{\rm (JgJ)}{\rm (JJ)} = (1.13 \pm 0.12 \pm 0.11)
\%\ .
 \end{equation}
{}Furthermore, the jets are found to be back--to--back correlated
in $\phi^{{\rm jet}1}-\phi^{{\rm jet}2}$.

In {\sc Phojet} using SCR model as described in Section III 
we find with the D0 trigger
 \begin{equation}
 R_{\rm JgJ-PHOJET-D0} = \frac{\rm (JgJ)}{\rm (JJ)} = 0.43  \%\ .
 \end{equation}
Here 0.1\% background JgJ events with only an accidental gap was
subtracted, this background was determined in a run without the
use of SCR.

With the CDF trigger we find
 \begin{equation}
 R_{\rm JgJ-PHOJET-CDF} = \frac{\rm (JgJ)}{\rm (JJ)} = 0.50  \%
 \end{equation}
where 0.5\% background JgJ events had to be subtracted.
\begin{figure}[thb] \centering
\begin{center}
\unitlength1mm
\begin{picture}(135,56)
\put(0,0){\epsfig{figure=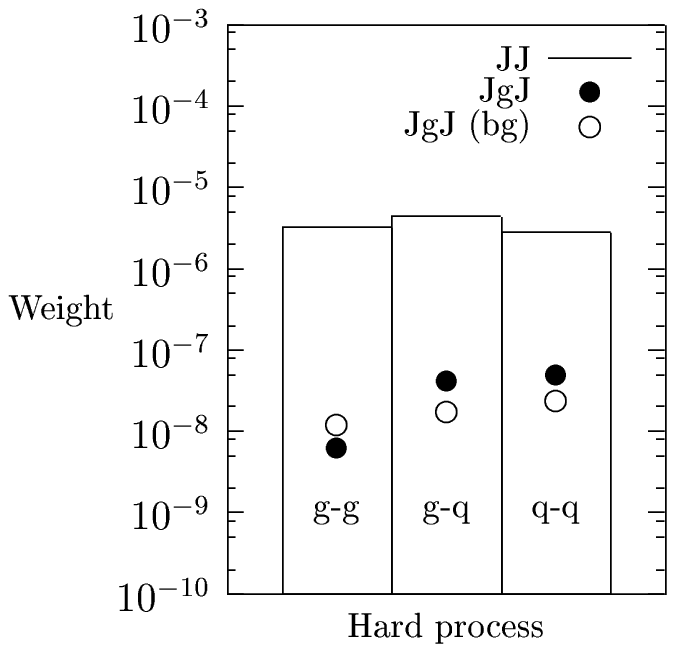,width=6.0cm}}
\put(2,3){a)}
\put(65,0){\epsfig{figure=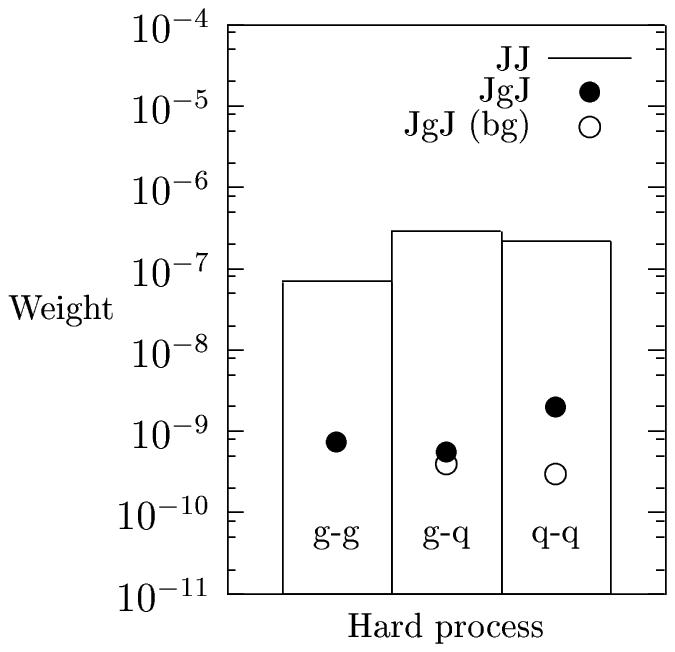,width=6.0cm}}
\put(67,3){b)}
\end{picture}
\end{center}
\vspace*{-3mm}
\medskip
\caption{(a)
Monte Carlo predictions for the fractions of g--g, g--q
  and q--q hard scatterings in the JJ events (without gap trigger), JgJ
   events obtained with SCR and background (bg) JgJ events (obtained
      without SCR) for the CDF trigger.
(b) The fractions of g--g, g--q
  and q--q events for JJ events (without gap trigger), JgJ
   events obtained with SCR and background (bg) JgJ events (obtained
      without SCR) for the D0 trigger.
\label{wjetjj}
}
\end{figure}

In the {\sc Phojet} Monte Carlo  we can subdivide the hard scattering
events into g--g, g--q and q--q scatterings. In Fig.\ref{wjetjj}
we plot for the CDF and D0 triggers the fractions of g--g, g--q
and q--q events for JJ events (without gap trigger), JgJ  events
obtained with SCR and background JgJ events (obtained without
SCR). At both energies we find, that q--q scattering dominates
the JgJ events, but g--q and g--g scattering contributes also.
{}For the q--q events the fraction of JgJ events due to SCR
(background subtracted) is always smaller than 1\%  of the q--q
scatterings without gap. 
In principle we could calculate from this separately the gap
survival probabilities for q--q, g--q and g--g scatterings, with
the present statistics of JgJ events however, we prefer not yet to give
such numbers. In
JgJ events we find generally only one or two pairs of jets whereas
in  events without a large gap the average number of jets is
definitely larger than this.

\begin{figure}[thb] \centering
\begin{center}
\unitlength1mm
\begin{picture}(135,62)
\put(0,0){\epsfig{figure=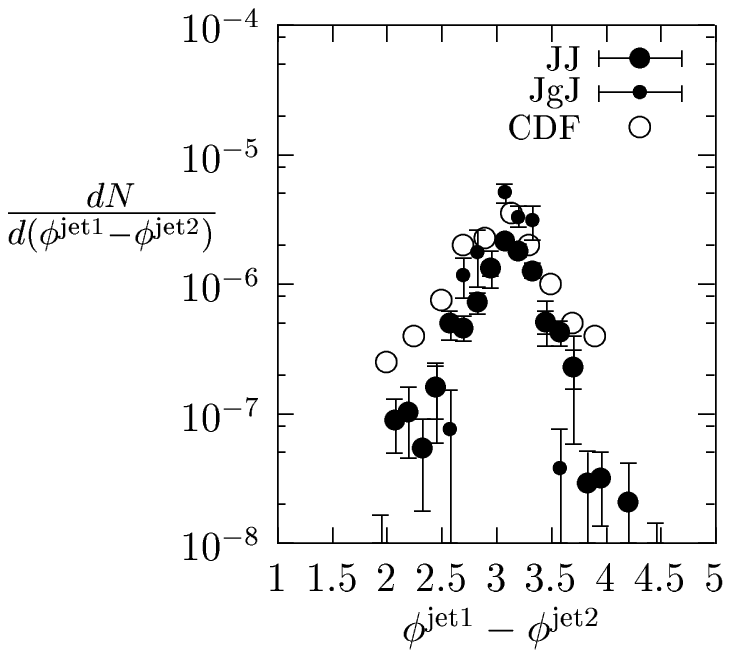,width=6.0cm}}
\put(2,3){a)}
\put(65,0){\epsfig{figure=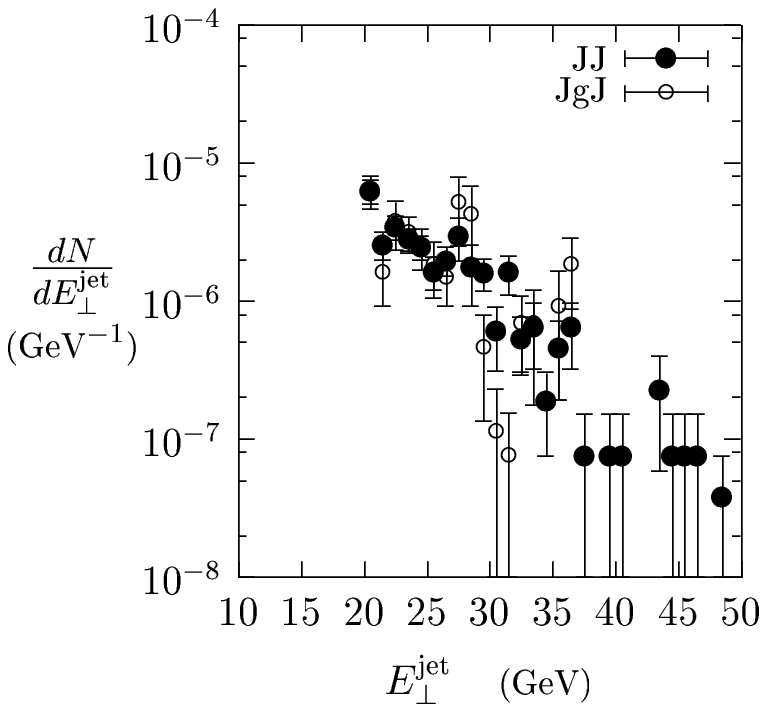,width=6.0cm}}
\put(67,3){b)}
\end{picture}
\end{center}
\vspace*{-3mm}
\medskip
\caption{(a)
The $\phi^{{\rm jet}1}-\phi^{{\rm jet}2}$
distributions corresponding to the CDF trigger \protect\cite{Abe98a}.
(b)
The $E_{\perp}^{\rm jet}$
distributions corresponding to the CDF trigger.
\label{fdndphij12}
}
\end{figure}

In Fig.~\ref{fdndphij12}.a we present the 
$\phi^{{\rm jet}1}-\phi^{{\rm jet}2}$ distribution 
calculated with {\sc Phojet} for the JJ and JgJ events for the CDF
trigger together with the corresponding distribution published
by CDF \cite{Abe98a}. All distributions are rather similar.
However, with bigger statistics we expect to see a more narrow
correlation in the JgJ events, in which further jets are much
less frequent than in the JJ events. In Fig.
\ref{fdndphij12}.b we present the calculated $E_{\perp}^{\rm jet}$
distributions. Within the statistics of the Monte Carlo runs we
do not find differences between the  distributions corresponding
to the JJ and JgJ events.

%
%
\begin{figure}[thb] \centering
\epsfig{figure=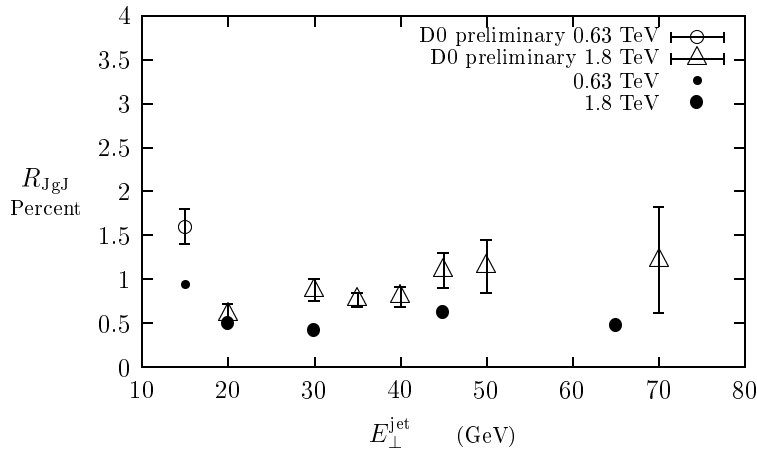,width=10.0cm}
\medskip
\caption{
The change of $R_{\rm JgJ}$ with the $E_{\perp}^{\rm jet}$, Preliminary
data from the D0 Collaboration \protect\cite{Abbott97}
are compared to the {\sc Phojet}
results obtained with SCR. 
\label{rjgj}
}
\end{figure}
%
The change of $R_{\rm JgJ}$ with the $E_{\perp}^{\rm jet}$ was studied
by the D0 Collaboration \cite{Abbott97}.
 A modest rise of the color singlet
 fraction with  $E_{\perp}^{\rm jet}$ was found. In Fig.~\ref{rjgj}
 we compare the {\sc Phojet} results on $R_{\rm JgJ}$ with these preliminary
 data. The {\sc Phojet} predictions exhibit a flat $E^{\rm jet}_{\perp}$ 
 dependence being still compatible with the data \footnote{Very recent
data of the CDF Collaboration \protect\cite{Abe98a} on a similar ratio 
show also a flat $E^{\rm jet}_{\perp}$ dependence}.
 The D0 Collaboration \cite{Abbott97}
 also found $R_{\rm JgJ}$ at $\sqrt
 s$ = 630 GeV to be a factor 2.6$\pm $ 0.6 larger than at $\sqrt
 s$ = 1.8 TeV. The ratio  of $R_{\rm JgJ}$ at these energies calculated
with {\sc Phojet} is consistent with the data, see  Fig.~\ref{rjgj}.

\section{Conclusions and summary\label{summary}}

 The processes implemented in {\sc Phojet} allow to study hard and
 soft diffraction (e.g.\ $\Pom$--$p$, $\Pom$--$\gamma$ and $\Pom$--$\Pom$
 collisions) in many channels. We hope that this tool and
 forthcoming data on hard and soft single diffraction dissociation
 and central diffraction from TEVATRON and HERA 
 help to answer important questions like:
 (i) Is soft color reconnection the correct mechanism to
 describe color singlet exchange processes between jets? Could
 this mechanism be responsible for other features of diffractive
 processes as well? (ii) Can hard diffraction
 consistently be described by pomeron structure functions? What
 is the low-$x$ behaviour of the pomeron structure function?
 (iii) Are there multiple soft and multiple hard collisions in
 diffraction like in $p$--$p$, $p$--$\gamma$ or $\gamma$--$\gamma$
 interactions?
 (vi) Does a super-hard component of the pomeron exist 
(e.g. can the data interpreted with a direct pomeron-quark coupling)?

Single-inclusive jet pseudorapidity distributions show only a small
sensitivity to a possible super-hard structure of the pomeron
(see Fig.~\ref{pt100jpopo}.b).
By contrast it is possible to use the jet transverse momentum
distribution to explore a super-hard pomeron structure. Furthermore, 
in the model, diffractive events containing jets produced by direct
pomeron-parton scattering exhibit much less soft hadronic background
than other JJg events. This soft underlying event feature might
allow for a crucial test of the models on a super-hard pomeron.
However, without using pomeron PDFs tuned to the latest HERA data reliable
predictions for TEVATRON energies cannot be obtained.

All features of the JgJ events observed at TEVATRON so far are
qualitatively well described in the Monte Carlo implementation of
the SCR model \cite{Eboli97a}.
However, first comparisons with data indicate that the ratio JgJ/JJ obtained
with a simple SCR model is too small. Further
investigations, higher statistics in the Monte Carlo events  and
more precise data are required to draw definite conclusions.

\noindent
{\bf Acknowledgments}\\
The authors are grateful to  S.~Roesler for many
discussions. One author (R.E.) thanks T.K.~Gaisser for helpful comments
and remarks.
The work of R.E.\ is supported by the
U.S.\ Department of Energy under Grant DE-FG02-91ER40626.



\end{document}